# *Attilio Sacripanti*

# Biomechanical Optimization of Judo:

# A sharp Coaching tool

(Practical Application and Scientific background)

## *Abstract*

In this paper, for the first time, there is comprehensively tackling the problem of biomechanical optimization of a sport of situation such as judo.
Starting from the optimization of more "simple" sports, optimization of this kind of complex sports is grounded on a general physics tool such as the analysis of variation.
The objective function is divided for static and dynamic situation of Athletes' couple, and it is proposed also a sort of dynamic programming problem "Strategic Optimization".
A dynamic programming problem is an optimization problem in which decisions have to be taken sequentially over several time periods "linked" in some fashion.
 A strategy for a dynamic programming problem is just a contingency plan. i.e., a plan that specifies what is to be done at each stage as a function of all that has transpired up to that point.
It is possible to demonstrate, under some conditions, that a Markovian optimal strategy is an optimal strategy for the dynamic programming problem under examination.
At last we try to approach the optimization of couple of athletes motion.
Shifting's Trajectories study is a very hard and complex work, because falls among fractal, self-similar trajectories produced by Chaotic (irregular) Dynamics.
It is possible to Optimize Athletes' motion planning?
Motion planning could be defined as the problem of finding a collision-free trajectory
from the start configuration to the goal configuration.
It is possible to treat motion as an optimization problem, to search the trajectory that drive adversary into a final *broken symmetry position*, where it is possible to *collide* and apply one of the two throwing tools ( *Couple or Lever*).

# *Attilio Sacripanti*

## Biomechanical Optimization of Judo: A sharp Coaching tool

### (Practical Application and Scientific background)

### 1. Introduction

Optimization is a powerful and flexible tool applied in modern Engineering and Economics to "optimize" some situations, as: income, structures and so on.
Optimization theory is central to any problem involving decision making, both in engineering or in economics. The task of decision making entails choosing among various alternatives.
This choice is governed by our desire to make the "best" decision. The measure of goodness of the alternatives is described by an *objective function*.
Optimization theory and methods deal with selecting the best alternative in the sense of the given objective function.
In very simply mathematical words, the goal of optimization is to build an *objective function* (linear or nonlinear) describing the process and to find a useful extreme (Maximum or Minimum) of this function.
Ancestor of this approach in Physics is the *Principle of Least Action*, or in better and more correct way the *Principle of the Stationary Action,* that is a variational principle that, when applied to the action of a mechanical system, can be used for example to obtain : the equations of motion of the system, the shortest path between two points , the best way of minimum of energy consumption, the minimum time trip, etc.
Many physical–mathematical experts spoke about this principle in old times: Fermat, Euler, Maupertuis, Hamilton, Leibnitz, and so on.
Credit for its most general formulation is commonly given to Pierre Louis Moreau de Maupertuis, who asserted: "the laws of movement and of rest deduced from this principle being precisely the same as those : observed in nature, we can admire the application of it to all phenomena:  movement  of animals, vegetative growth of plants ... are only its consequences; and the spectacle of the universe becomes so much the grander, so much more beautiful, the worthier of its Author, when one knows that a small number of laws, most wisely established, suffice for all movements." [1]
In application to physics, Maupertius suggested that the quantity called "Action" to be minimized was the product of the duration (time) of movement within a system by the "vis viva", which is the integral of twice what we call the kinetic energy $K$ of the system, in formulas:

$$\delta A = \delta \int_{t}^{t} 2K(t)dt \qquad (1)$$

When we speak about best decision or best way to minimize energy consumption, these two extremals give us the amplitude of application of Optimization theory.
The calculus of variations is concerned with problems in which a function is determined by a stationary variational principle. In its simplest form, the problem is to find a function $y(x)$ with specified values at end-points $x_0$, $x_1$ such that the integral $J = \int_{x_0}^{x_1} f(x, y, y')dx$ (2) is stationary.

Normally apply the calculus of variation to sport means to find a value to minimize energy, or muscles effort, or the right angle to obtain the optimum performance in sport. [2]

## 2. Optimization in Sports

In term of Biomechanics, sports can be classified as: Cyclic, Acyclic, Cyclic - Acyclic Alternate and Situation Sports (dual and team) [3].
For cyclic and acyclic sports optimization is complex but possible, while for situation sports Optimization is very complex and sometime impossible.
For example, it is worldwide known the verbal expression of the principle of stationary action for the human locomotion (which is a cyclic motion) that can be express by the following words:"necessary and sufficient condition to have a locomotion is that the algebraic sum of all joint's angles in the human body must be zero".
When optimization is applied, to simple cyclic sports like cycling and swimming, the *objective function* ( in theoretical optimization) is to build the best performance, more often in practical optimization, for cycling, this *objective function* is connected to the maximization of the rational muscular effort finding the most effective angle of application of force on the cycle.
The best performance in practical optimization could be also applied to the minimization of external parasitic forces, like drag, for example in swimming, but for these simple biomechanics cyclic sports also both goals can be prosecuted when the performance is an high performance level.
In the following figures it is shown the optimization for cyclic sports like Cycling and Swimming .Fig.1,4

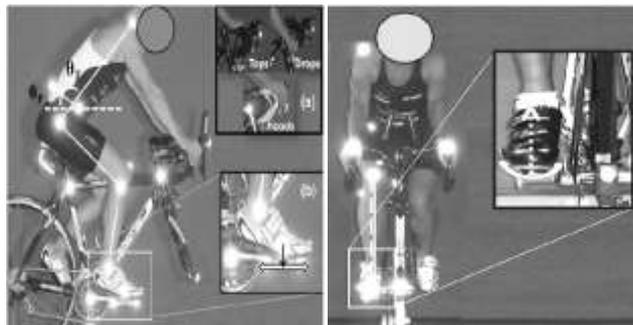

Fig.1,2 Cycling Optimization finding the best angles for athlete muscles [4]

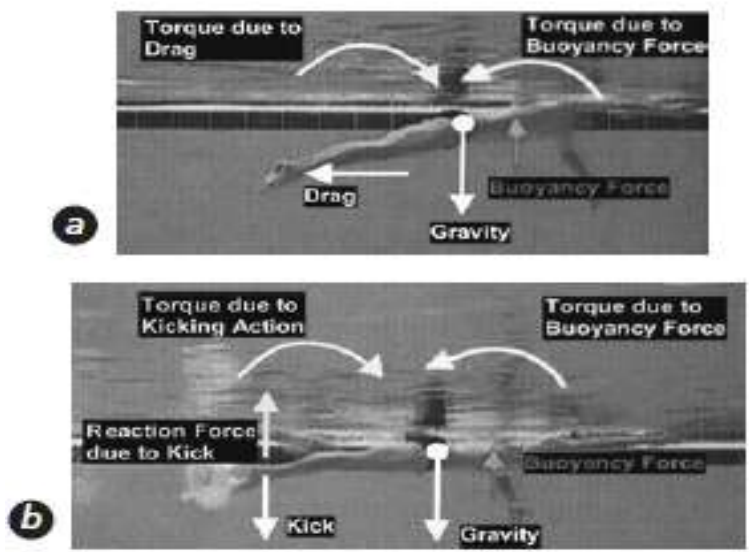

Fig.3,4 Swimming Optimization against Drag [5]

In the previous sports, for coaches, practical Optimization means: to find the "optimum" angles to let muscles working in best way, or to work efficiently with minimum effort against Drag.

From the previous short exempla it is easy to understand that in Sport, optimization is evaluated as *objective function* in theoretical way, but applied in practical way or maximizing the "internal" capabilities of the athletes or minimizing the "external" parasitic forces acting against the performance; where "internal" and "external" depend on the definition of the "athlete system".
Optimization [6][7] is a very flexible tool that comes from physics variational calculus and widely applied in modern engineering and economics, theoretically it is performed, finding the algorithm to minimizing for example energy or maximizing income, in Sport Science Performance Optimization is connected to biomechanics, ergonomics, nutrition, etc.

### 3. Optimizations in judo

Different problem is the Optimization of a complex Sport belonging to Situation Sports group. In such group pattern are everywhere changing, motion is very complex and situations happen only with statistical frequency.
Then the *objective function* is very hard to find or to build and optimization is very difficult.
Judo is a Dual Situation Sport and its global optimization is very complex and sometime not affordable.
But there is a way to overcome the structural difficulties shown before, along the line of the Cyclic Sports Optimization, it is possible to apply Optimization by differential method, or in easy words: dividing Judo, step by step, in appropriately selected subsets.
It is interesting to understand that these subsets chosen on the basis of the dynamics of movement of Couple of Athletes System are able to give us information about optimization of the "Attacker" that is a component of Couple of Athletes System then associating to each subset a specific *objective function* it is possible to optimize all techniques to the best performance.
For Coaches the best way to perform practical Judo optimization is to apply, qualitative Biomechanics, in two situations and three areas:

**Situations**

*Teaching lessons* - Couple condition Static.
*Competition* – Couple condition Dynamic.

**Areas**

Couple Statics ⟶ OF = ***Minimum*** Energy
Couple Dynamics ⟶ OF = ***Maximum*** Effectiveness

Couple Long Development Dynamics ⟶ OF = Strategic (Overall ***Minimum*** Energy) Effectiveness

### 4. Situation 1: Teaching Lessons - Statics - Minimum Energy Optimization

The most feasible and easy approach to this complex sport is to analyze first the Couple System in static situation for optimization this means still Athletes and shifting velocity of Couple zero.
In Static fixed Situation if Coaches analyze Judo Interaction, (throws, old-down, joint-break and choking).
Biomechanics let to optimize (as suggested Kano) on the basis of Minimization of Energy expenditure. [8].
Osaekomi waza, (old down), Kansetzu Waza, (Joint break), Shime Waza (choking) are performed, in accord with physics laws, with less energy consumption [9].
If Coaches analyze Classical throws classified in five groups and arranged into the Go Kyo (five lessons) at light of Biomechanics find that they can be grouped in only two classes:

1. Lever System    More Energetically expensive
2. Couple System Less Energetically expensive [10]

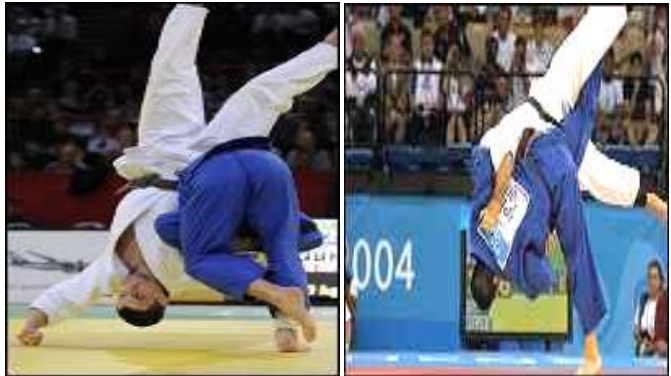
Fig.5-6 Most utilized throws in competition: Seoi ( Lever) , Uchi Mata ( Couple)

The static action of throwing techniques goes through some specific phases: Unbalance, Positioning and Throw, or in Japanese tradition ( Kuzhushi, Tsukuri, Kake) the first two phases can change during competition, generally speaking all classical throws in judo need three types of trajectories to shorten the distance between athletes ( rectilinear, inward rotation , outward rotation)

*Trajectories (rectilinear or inward rotation)*
Optimizing short trajectories
The short trajectories that shorten distances between athletes can be optimized at light of calculus of variation.
In this specific situation it is possible to apply the Euler/ Fermat form of the principle of the Stationary Action. Euler /Fermat principle:
*The curve described by Athletes' Body to shorten distance between two points A,B is the curve (among all the possible paths connecting A,B) that minimizes the momentum mv , Athlete travel between these two points along the path of shortest time.*

$$\partial S = m \int_A^B v ds = \min. \qquad (3)$$

This form of principle more understandable in term of mechanical movement will satisfy optimization trajectories during shortening distance between athletes.
In such way the three shorten distance trajectories, called General Action Invariants (GAI), will be minimized and the work performed by Athletes' Center of Mass ( COM) shall be minimum.

*Lift-up*
The action of lifting is able to facilitate throwing action by reducing the friction between the feet and Tatami, it is of great help both in the techniques of Couple in the frontal plane and with inward rotation, practically useless for Couple techniques with straight approach ( es. O Soto gari, ecc.).

*Almost plastic collision of extended soft bodies*
All throws, after the shortening of mutual distance, need a collision between athletes' bodies.
In the case of Couple Group, both straight trajectory attack ( O Soto Gari ) and inward rotation attack ( Harai Goshi, Uchi Mata) , the greater surface contact collision depends upon the chest that must apply upper force of Couple to opponent's chest , collision of a smaller area is the responsibility of the leg that applies lower Couple force to the opponents' leg.
So in the case of Couple at the end of the previous GAI trajectories occurs a simultaneous contact/collision that can be considered almost plastic, because the two athletes are closely related, and Uke tends to rotate around his fixed Center of Mass and more is the simultaneous application of the forces and longer equivalent their intensity, more the actual motion will approach the theoretical one's. The mechanics is obviously different from the Lever tool application.
 In the Lever case Uke, subjected to the Tori torque applied, tends to rotate not around its Center of Mass, as in Couple Group, but around the stopping point (fulcrum) and his Center of Mass translates in space . All this is already analyzed in some old papers, among others [11],[12].[13].

   *Flight Paths*
In accord with the Kano approach astoundingly Classical form of judo throws, careful chosen by Kano, are optimized: during their flight paths after the application of throwing tools (Couple or Lever) Uke' body goes through the geodesics of two symmetries: cylindrical or spherical.
It is well known that geodesics are the shortest paths between two points on a specific curve, result of the calculus of variation  [14] ( See Appendix 1)  In Formulas in our metric space considering the Cylinder and Sphere as basic symmetry figures for Tori the two minimizing geodetics satisfy the relationship:

$$dist(c(t_1), c(t_2)) = |t_1 - t_2| \qquad (4)$$

Fig 7,8,9,10,11.

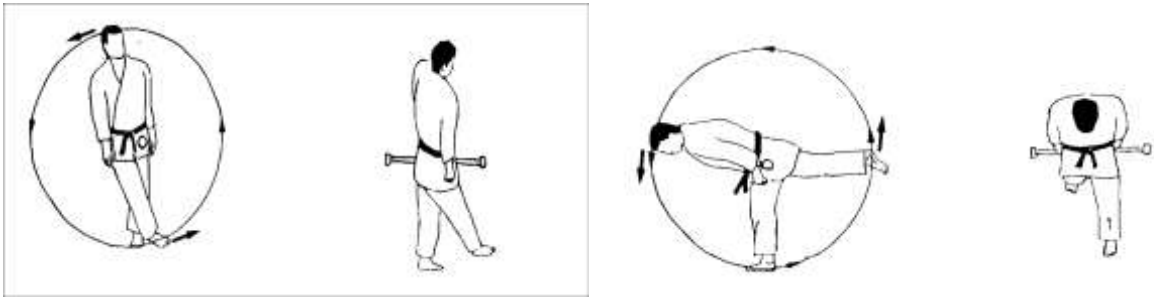

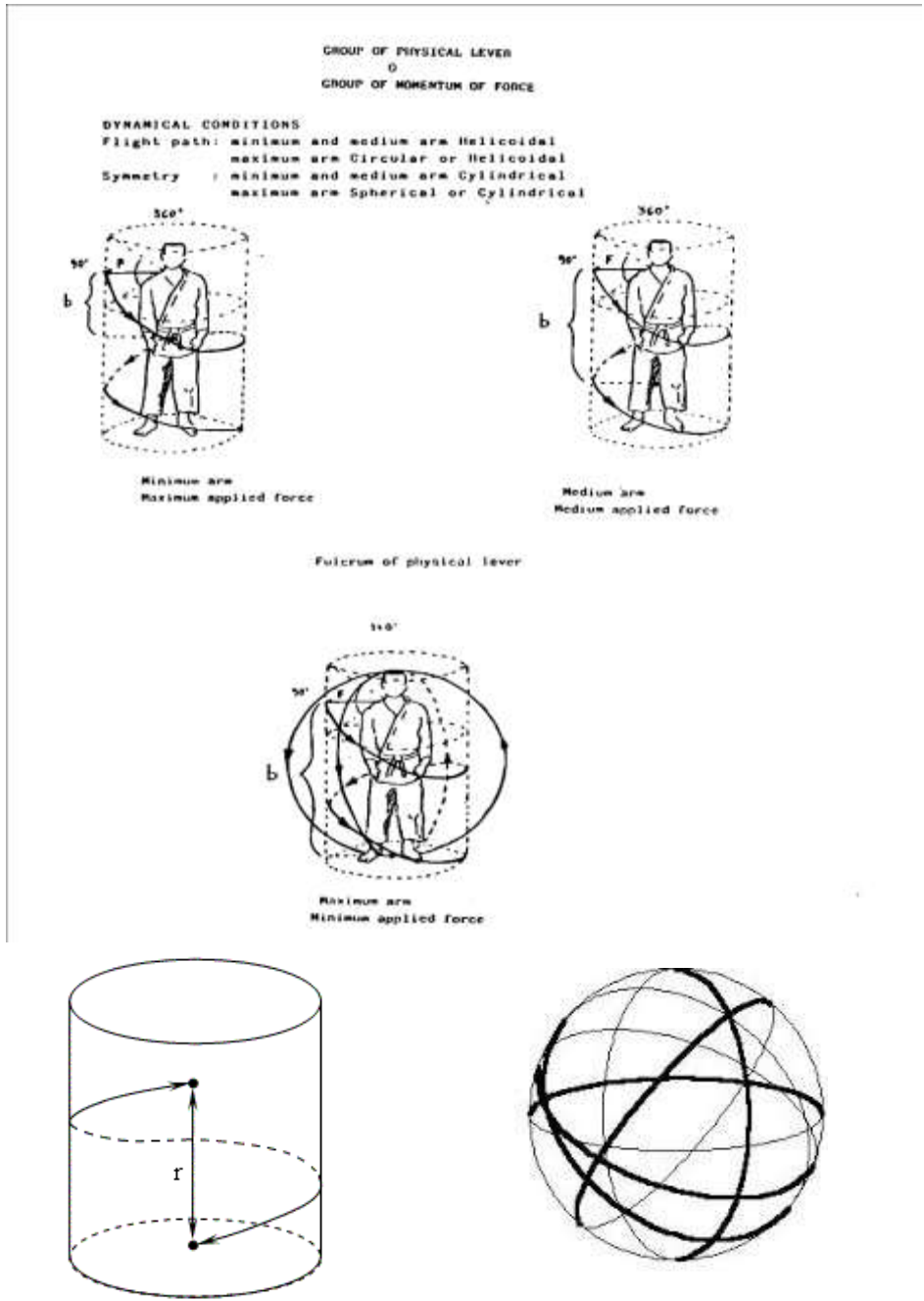

**Fig.7,8,9,10,11 geodetics of symmetries of throwing actions [15]**
**(see Appendix 1)**

From the previous short analysis it is possible to underline:

In the Couple Group Uke's Center of Mass (COM) turns around himself, all these throws are, theoretically speaking, gravity independent, less expensive, **Fully Optimized.**

In the Lever group Uke's Center of Mass (COM) shifts in space, throws are gravity and friction dependent, more expensive, *not fully optimized*.

However they can be optimized *changing length at the arm of the lever,* with **objective function** Minimization of Energy.

For example from standing Seoi [Ippon Seoi Nage], to Kneeling Seoi [Seoi Otoshi], till to Drop Seoi [Suwari Seoi], passage that is only the Optimization of the same Lever throw.

But in Japanese vision, the previous optimization of one throw, they are three different throwing techniques!

## 5.     Competition: Dynamics- Maximum Effectiveness Optimization

If Coaches analyze dynamical situations (competition), Optimization grounded on the *objective function* Minimization of Energy Expenditure is a necessary condition but not sufficient.

In fact it makes appropriate to expand the Optimization goal with a wider *objective function* not simply connected to a minimum of energy, but which considers also different energy consumption in function of Maximum Effectiveness. Maximum Effectiveness means not only that minimization of energy is working like static situation, Classical application of Kano's Throws is possible in competition if there is a big gap between athletes. Fig 12,13

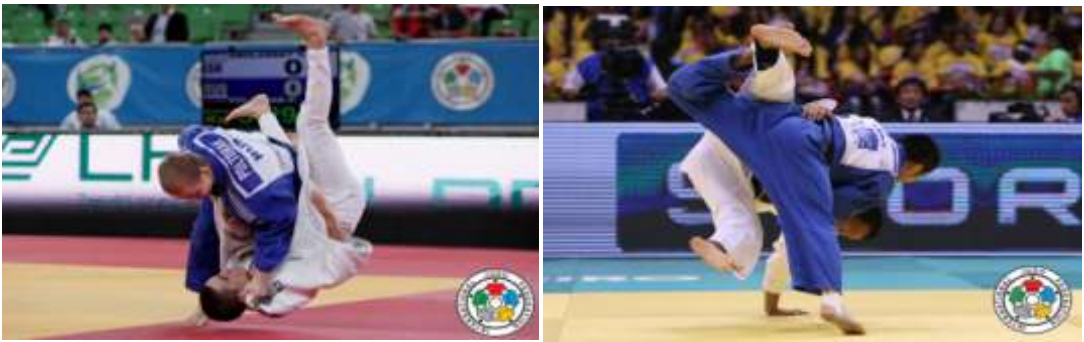

*Fig 12,13 Optimization of energy in classical throws*

Coaches must consider also not full energetically convenient actions that could be Optimum as very effective like the following action shown in the next fig. 14, 19

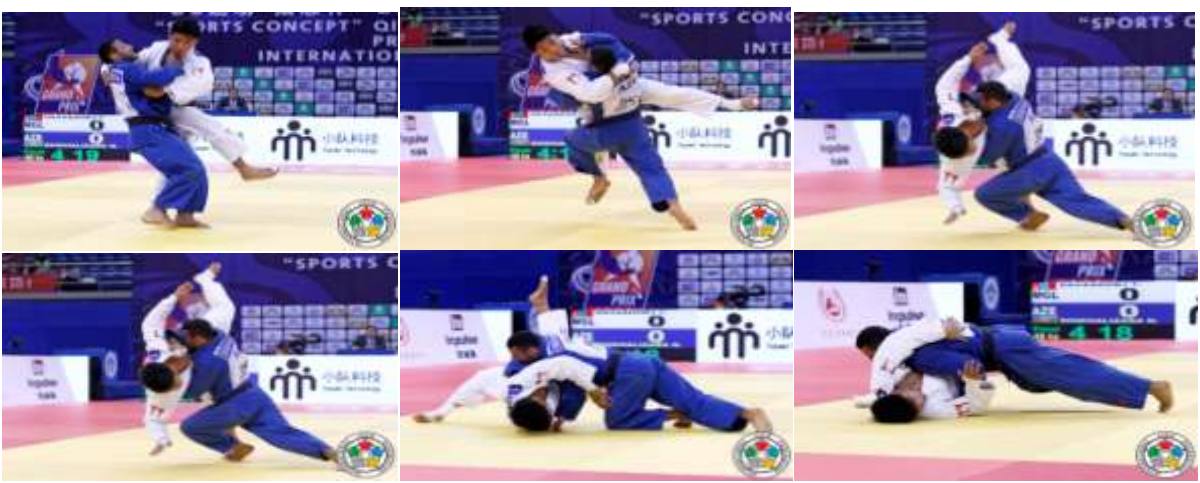

*Figg. 14-19 Effectiveness Optimization, expensive but effective throwing action*

Coaches must optimize and replaces Kano's theoretical unbalance concept or application with the practical exploitation of the "breaking symmetry" [12]

The "Breaking Symmetry" concept is a more subtle and practical way to optimize the application of Judo throws in high dynamic situations.

In Static situation Teacher unbalances Uke's body and applies the tool to throw, this preliminary unbalancing action is performed to simplify the throwing action using the gravity force and the unstable equilibrium of Uke as complementary tools.

In competition it is very difficult if not impossible to simply unbalance the Adversary that defends himself hard, Tori to attack effectively must prevent fast avoidance, then forcing Uke's body to turn or bend practically he causes the center of mass moves inside the body.

When center of mass shifts to one side stability increases, mobility decreases and Tori must apply, on the right (more stabilized) side, Couple or Lever tool to throw simultaneously to a collision of bodies.

The previous one is the right and optimized sequence in competition to practically throw the adversary.

Practical Effectiveness Optimization passes through the use of specific complementary tools that are meant to improve the efficiency of throwing techniques although at the expense of energy.

These complementary tools are essentially based on the exploitation of the weak points of the human body from the biomechanical point of view or on the energy specific application

Tools in Throws Practical Effectiveness Optimization are

1. Crosswise attack directions.
2. Technical variations with energy saving changing mechanics as Spinning Seoi.
3. Changing Couple in Lever and vice versa
4. Rotation in transverse plane,
5. Pure Rotational Application
6. Use of Chaotic techniques. [16]

## 6. Competition: Some Problems

Judo Competition in Biomechanical and physical terms is a complex, nonlinear system with fractal self-similar trajectories [8]

During Interaction open skill evolves continuously in infinitive specific technical tactical solution. More often, in such fast changing situations, for Coaches it is not possible plan any practical optimization.

In the following figures there is one exemplum of fast changing situations almost connected and grounded on the advanced personal skill of high level competitors.

Judo as Open Skills Sports presents very complex situations that evolve continuously, from standing to lying fighting on the ground.

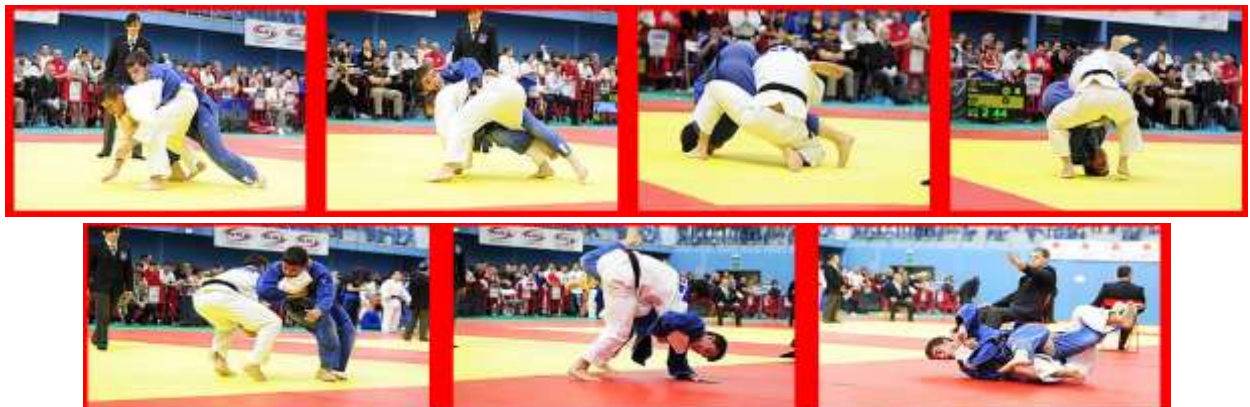

*Fig 20-26 Ne Waza and Nage Waza continuous changing Skills combined*

# 7. Competition: A more wide Optimization Concept- Strategic Optimization

Strategic Optimization is theoretically connected both to the time of a *whole contest* and to a *sequence of contests*. In practice for Coaches it could be based on the minimum mechanical work made up of throws' right choice, specific tactical tools to enhance throws effectiveness, opportune ne waza's right use, locomotion, defensive work, and active recovery among contests, all evaluated for example as $O_2$ consumption.

Theoretically speaking, Strategic Optimization of a judo contest or sequence of contests is defined a *dynamic programming problem.*

*A dynamic programming problem* is an optimization problem in which decisions have to be taken sequentially over several time periods "linked" in some fashion.

A *strategy* for a dynamic programming problem is just a *contingency plan.* i.e., a plan that specifies what is to be done at each stage as a function of all that has transpired up to that point.

It is possible to demonstrate, under some conditions, that a Markovian optimal strategy is an optimal strategy for the *dynamic programming problem* under examination. [17]

In more easy words, if we build an Heuristic Energy Equation of the contest by a Constructive Algorithm. Which it is an algorithm that constructs by iterations the final solution by building upon a partial (incomplete) solution.

it is possible to find an optimal solution strategy for the contest by means of equation like this.

$$\overline{E(O_2)} \cong (\sum_{j,c} A_{j,c}^{o_2} + \sum_{k=i} mgt_a \overline{v_{k=i}^{o_2}}) + \sum_{n,i} D_{n,i}^{o_2} + \sum_n G_n^{o_2} + \mu v^{o_2} x^{o_2} + WAR^{o_2} - \sum_s \frac{t_s}{\tau} \overline{E(O_2)}$$

$A_{j,c}^{o_2} = Tachi\ Waza\ attacks\ as\ O_2\ (j = classics; c = chaotics)$

$mgt_a \overline{v_{k=i}^{o_2}} = Work\ of\ Tachi\ Waza\ attacks\ as\ O_2$

$G_n^{o_2} = Ground\ Works\ as\ O_2$

$D_{n,i}^{o_2} = Defensive\ Works\ as\ O_2\ both\ for\ Tachi\ and\ Ne\ waza$

$\mu v^{o_2} x^{o_2} = Work\ of\ couple\ motion\ as\ O_2$

$WAR^{o_2} = Work\ for\ Active\ \mathrm{Re}covery$

$t_a = time\ of\ attack$

$t_s = time\ of\ stop\ (Matte)$

$\tau = regular\ fight\ time$

Obviously, with this **Objective Function** the optimization is connected to the minimum in mean of the singular contributions less the Tachi Waza attacks that could be also optimized on the maximum effectiveness Optimization as already seen.

Practically for coaches it is not possible to find a minimum point objectively by measurements. The best way to obtain the strategic optimization in real competition is by visual observation and expert vision (knowing the mean experimental measurements of athletes consumption) and then to consider a reasonable range of conduct in competition in which energy consumption will be acceptably low.

# 8. Scientific Study of complex System: Shifting's Paths Optimization

## 8.1 Macroscopic Level : Locomotion

### *Locomotion   (Ayumi Ashi, Tsugi Ashi)*

The Judo locomotion in competition (Aruki Kata) can be divided in two groups Ayumi Ashi (normal locomotion) and Tsugi Ashi formally (foot follows foot).
Human walking biomechanics is a very complex branch of the Kinesiology, the actual knowledge on walking mechanics came from a lot of researches, the first ones at the start of the previous millennium. As we demonstrate in the book "Biomeccanica del Judo":  Tsugi Ashi is energetically convenient respect to Ayumi Ashi, and a bit more stable respect to external perturbation of push/pull forces.
Walking is a cyclic activity in which one stride follows another in a continuous pattern.
We define a walking stride as being from touchdown of one foot to the next touchdown of the same foot, or from toe-off to toe-off.
 In walking, there is a single-support phase, when one foot is on the ground, and a double-support phase, when both are. The single-support phase starts with toe-off of one foot and the double-support phase starts with touchdown of the same foot.
The coordinated human walking movements are generated not from an explicit representation of the precise trajectories of each anatomical segment as in bipedal robotics but by dynamic interactions between the nervous system, the musculoskeletal system, and the environment.
Different types of movement exist and are associated with different types of command.
 (1) Voluntary movements are integrated at the cortical level and can be initiated without any external stimulus.
(2) Automatic movements are memorized strategies that are elicited by internal commands or external stimuli.
(3) Spinal reflexes are genetically programmed responses to external stimuli, modulated by superior centers.
A primary concern of the CNS is to maintain dynamic stability during locomotion in competition. Dynamic stability during locomotion is the control of COM within a changing base of support and requires effective proactive and reactive recovery response strategies when exposed to perturbations
The motion of COM is not a simple motion, for reasons of dynamic stability of the athlete, as it is possible to see in the next figures in the symmetry planes of the human body. [18]

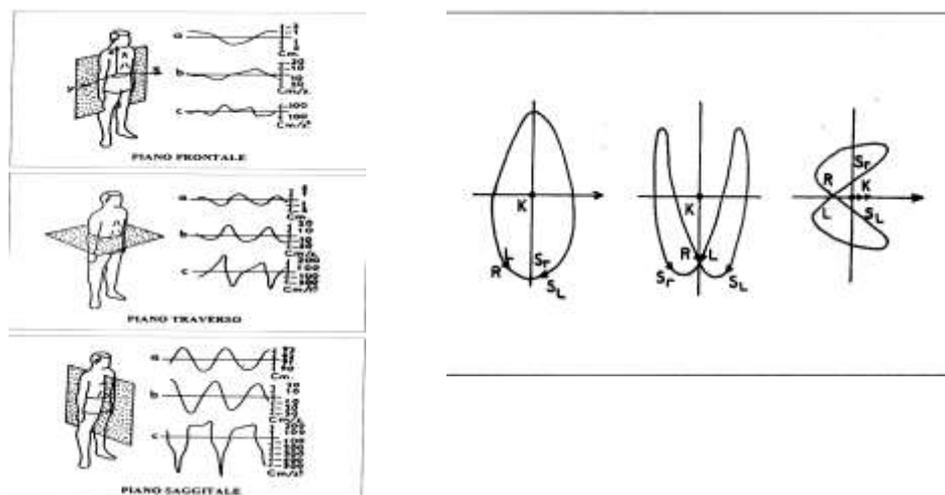

*Fig.27-28  Motion of COM in the three symmetry planes of human body and relative curves [18]*

The symmetry planes of the Human body are shown in the next figures in connection with the projections of the human body on the planes during the locomotion.

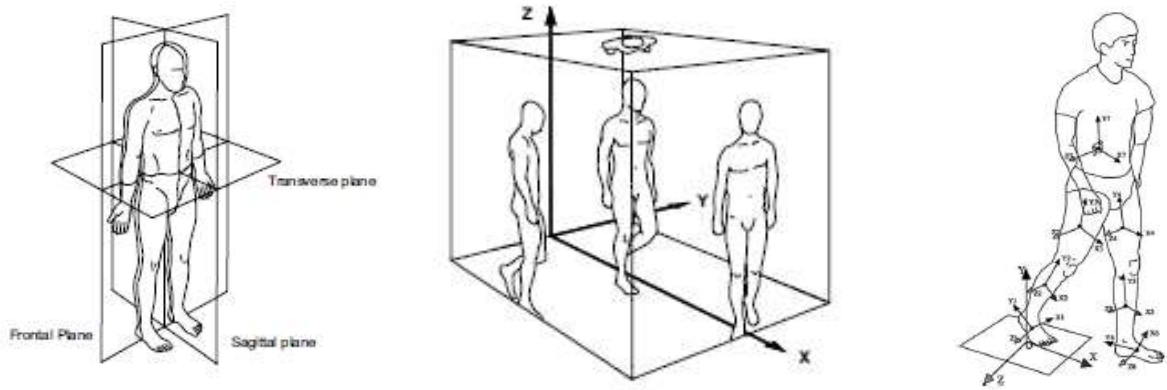

*Fig.30-31 Human body symmetry planes, projections of the body on the planes during locomotion and reference systems . [19][20]*

The curves presented in the previous figure 28 are the closed curves that the COM travels during a locomotion cycle. These curves that are representative of the motion of the COM during a gait cycle, in the interesting work of Minetti and co-workers [21] showed the basic equations for the construction of similar curves in connection with the symmetry of the closed loop and the Lissajous curves.

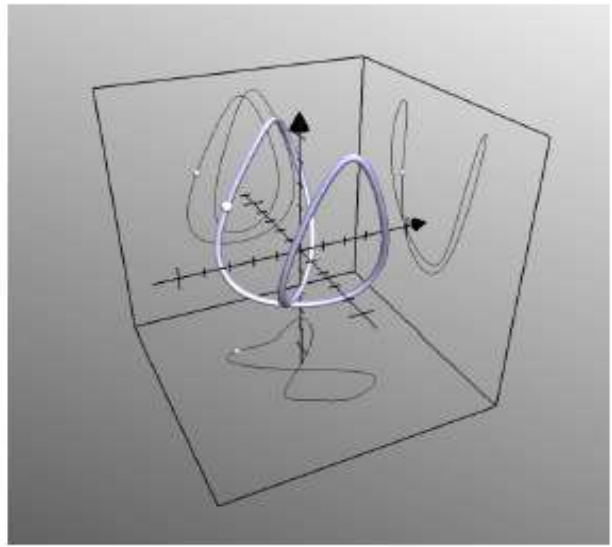

*Fig 32 Curves of the average trajectories of COM for a cycle, during walking at 5 Km/h [21]*

The equations are the following.

$$\bar{x}(t) = \bar{v}(t) + \sum_{i=1}^{6} \bar{c}_i^x \sin(2\pi f_i t + \bar{\phi}_i^x)$$

$$\bar{y}(t) = \bar{a}_0^y + \sum_{i=1}^{6} \bar{c}_i^y \sin(2\pi f_i t + \bar{\phi}_i^y) \qquad (5)$$

$$\bar{z}(t) = \sum_{i=1}^{6} \bar{c}_i^z \sin(2\pi f_i t + \bar{\phi}_i^z)$$

It is interesting to note that the previous curves apply only to the speed of 5km / h in fact, with the increase of speed of locomotion curves change shape while preserving similarity as shown in the next figures

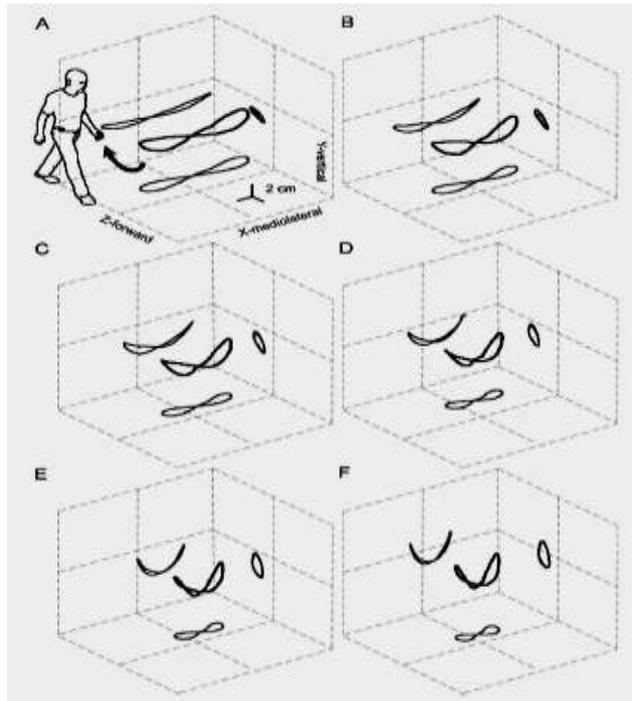

*Fig 33 Relationship between COM 3D trajectories and walking speed [22]*

*Locomotion Optimization*
It is interesting that, at macroscopic level, human locomotion is already the result of one Optimization process on the Joint angles, in fact locomotion is possible only under the condition that the sum of the joints' angles is zero.
However Optimization of Human locomotion is a complex two phase optimal problem, because real locomotion is a function of too many parameters.
By using optimal control methods it is possible simultaneously optimize the motion $x(t) = [q(t), v(t)]^T$ which consists of the positional variables $q(t)$ and the velocities $v(t)$ of the generalized coordinates of model. The torques at the actuated joints are described by $u(t)$.
Additional model parameters, such as spring damper constants or step length and velocity, are considered in the vector $p$. As there are two equations (one for single– and one for double–support) there are two–phase optimal control problem.
The complete optimal control problem can be expressed by the following **Objective Function**

$$\min \int_0^{t_n} L(x(t), u(t), \rho, t) dt + M(t_n) \qquad (6)$$

In which we have the Lagrangian which minimizes both the torques $u(t)$ applied to the system and the motion of the head, plus the Mayer term $M(t_n)$ that contains the impulse at the foot on touch-down[23]. During stance into locomotion the barefoot change its contact surface with the tatami as shown in the next figure.

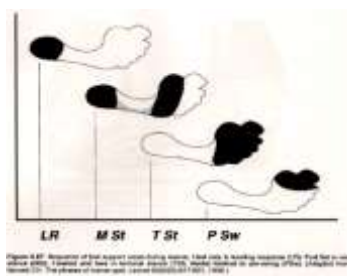

*Fig 34 Sequence of barefoot surface during stance [23]*

The mean interactions in judo competition are push and pull forces applied into the couple of athletes systems.
Few studies are been applied to this special part of the competition the response of the human body at such kind of perturbation during walking.
During maintenance of postural equilibrium, COM is kept within the base of support by activating appropriate muscles that move the Center of Pressure (COP). During locomotion, the same principle for the control of COM applies, with one important difference: foot placement at the end of each swing phase provides the primary method of moving COP in the sagittal and frontal planes.
The results from a recent study on the locomotion illustrate this clearly.
In this study, individuals were walking on a treadmill while unexpected mechanical perturbation (push) to the upper body in the frontal plane was applied during the two single support phases.
A push to the right when the left foot is in the single support phase produces abduction of the right swing limb and subsequent increase in step width.
In contrast, a push to the right given when the right foot is in the single support phase produces adduction of the left swing limb and subsequent decrease in step width see the following figure

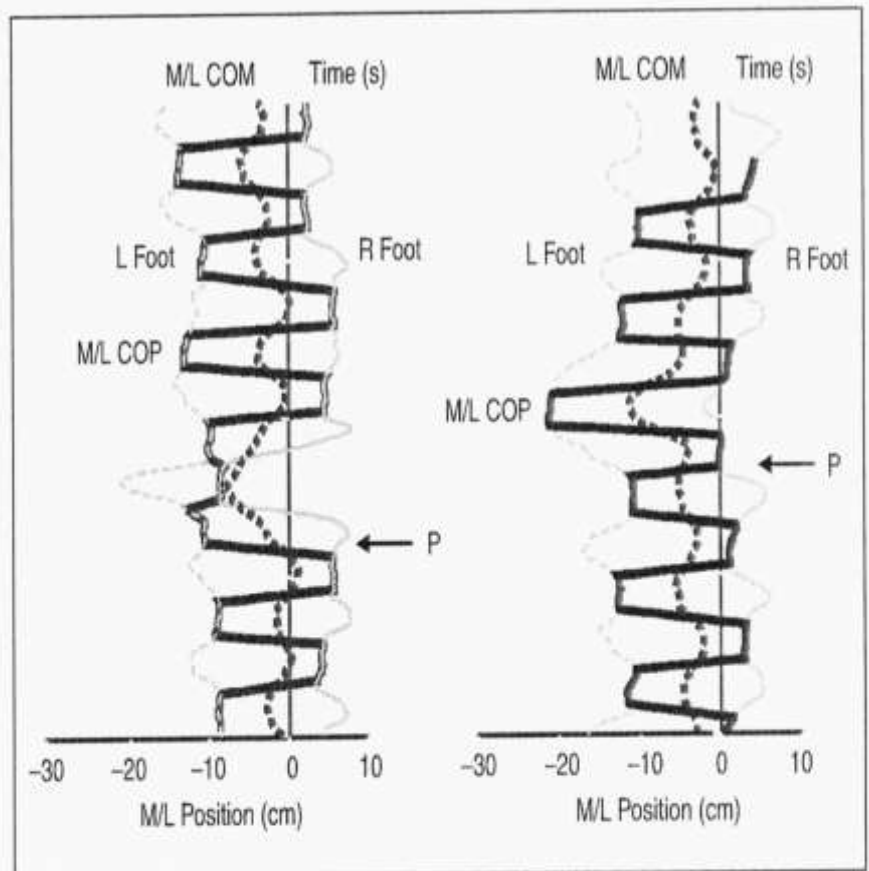

Fig. 3. COM and COP profiles before and following a push from the right at the shoulder level while the person was walking on a treadmill are shown. COP profiles are estimated from foot marker profiles. The left panel shows response when the left foot is on the ground: note how the right foot crosses over to be placed ahead of COM. The right panel shows response when the right foot is on the ground. Perturbation onset (P) is indicated by an arrow.

*Fig.35 Response reversal observed during medio-lateral perturbations, applied during Human locomotion [24]*

## 8.3. Mesoscopic Level approach.

*Center of Mass motion in still position*
In static equilibrium the CM (centre of Mass) projection and the COP (centre of pressure) would lie on the same plane, on the vertical line COP would coincide as proportional model with the projection of the CM on the ground. Both motions are similar but the COP motion is always larger than the CM projection motion. This can be illustrated in Biomechanics using a simple model, the inverted pendulum, Winter 1998, Pedotti 1987 for the anterior posterior balance.
The pendulum rotates around the ankle joint which we take as origin of the Cartesian system if we denote as F the force acting on foot by force plate at the point (-ζ, η) which is the COP.
The system is described in Newtonian approximation by the equations:

$$m\ddot{y} = F_y$$
$$m\ddot{z} = F_z - mg \quad (7)$$
$$I\ddot{\alpha} = \eta F_z + \zeta F_y - mgL\cos\alpha$$

The component $F_z$ is the same force as is obtained from the readings of the force transducers.
For a small deviation around the vertical z axis, we may replace cosα by y/L and in the first approximation we may also set, $F_z$= mg then the last equation will be:

$$y - \eta \approx \left(\frac{\zeta}{g} + \frac{I}{mgL}\right)\ddot{y} \quad (8)$$

After some easy manipulation and putting the equation in term of the angle π/2- α= θ we obtain:

$$\ddot{\alpha} - \left(\frac{mgL}{I}\right)\alpha = 0 \quad (9)$$

This equation of course describes an unstable situation, the inverted pendulum topples over.
However this classical procedure do not explain the Random Walks characteristics of quiet standing coordinates of the COP they can be explained by the equation of Hastings & Sugihara 1993 [26] that combines random walk with a friction term like a Langevin equation:

$$dx(t) = -rx(t)dt + dB(t) \quad (10)$$

Here dB is the uncorrelated noise with zero mean

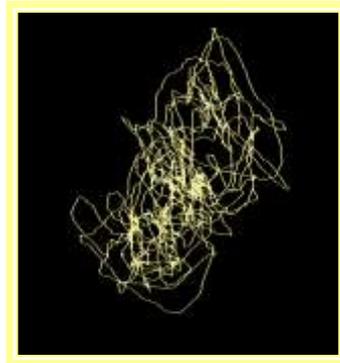

*Fig.36 Posturogram- Random Walk of the COP coordinates( for still athletes).*

*Multifractals in Human Gait*

Walking is a very complex voluntary activity, the typical pattern shown by the stride interval time series suggest particular neuromuscular mechanisms that can be mathematical modeled.
The fractal nature of the stride time series of human was incorporated into a dynamical model by Hausdorff using a stochastic model that was later extended by Askhenazi et others, so as to describe the analysis of the gait dynamics during aging.
The model was essentially a random walk on a Markov or short range correlated chain, where each node is a neural that fires an action potential with a particular intensity when interested by the random walker.
This mechanism generates a fractal process, with multifractals aspects in which the Holder time dependent exponent depends parametrically on the range of the random walker's step size.
The multifractal gait analysis is also used to study the fractal dynamics of body motion for patients with special aging problems or diseases, like Parkinson or post-stroke hemiplegic.
In the next figures we can see.
The variation of the time dependent Holder exponent, with the walker step size for free pace and metronome pace at different speed. [27]

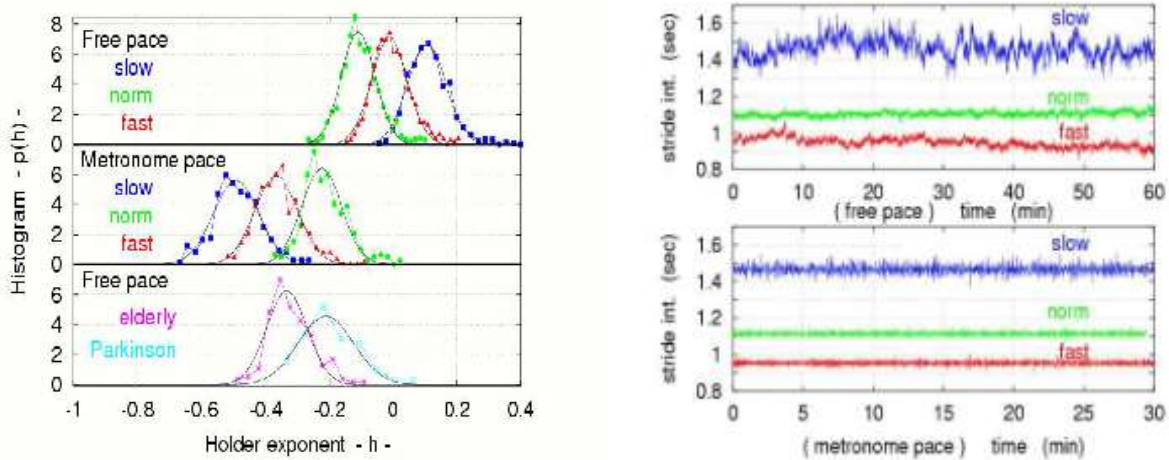

*Fig.37, 38, different time series produced by free pace and metronome pace at different speed.*

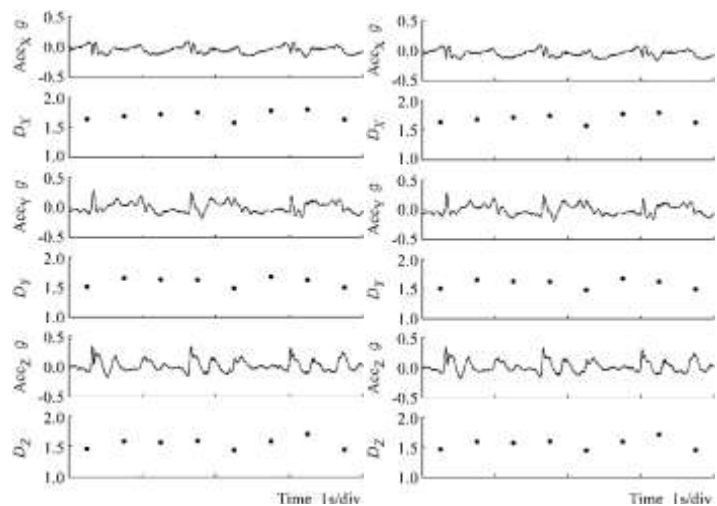

*Fig.39, 40 Relative acceleration signals and the related fractal values in post stroke and Parkinson patients.*

If we analyze the shifting paths in competition they are not simple lines but the true paths is affected by standard error produced by the deviation of center of pressure during gait.
The mean standard deviation for center-of-pressure paths during barefoot normal walking for judoka is shown as example in the next figure.

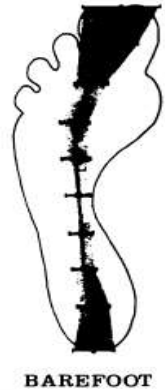

*Fig.41 Standard deviation of barefoot during normal locomotion ( Ayiumi Ashi)*

Then the paths shown in the Japanese figures really are the most probable shifting paths in time , not considering the standard error present.
If the function f(q,t) gives us the most probable shifting path in time, which is singled out by the maximum probability points of the same function during its time evolution, we can draw the path, in the next figure it is shown the choice that is produced when the shifting paths are draw.

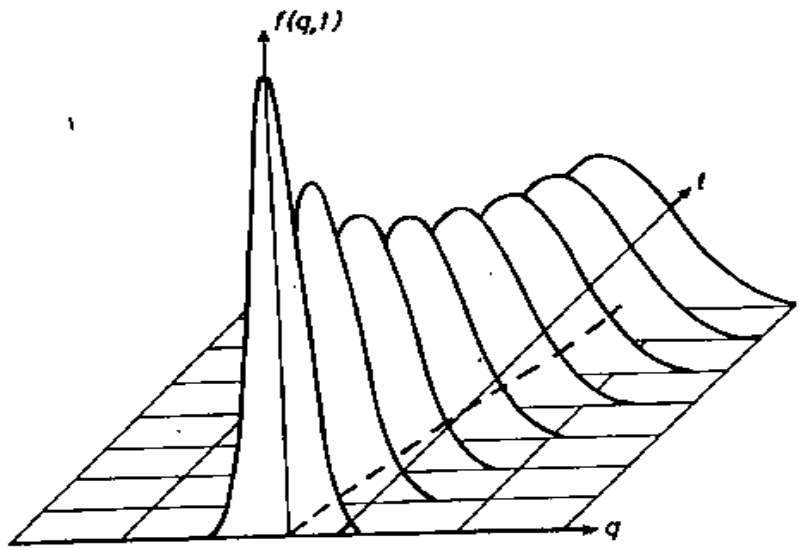

*Fig.42 Shifting paths as maximum probability trajectory*

## A micro/meso approach to Locomotion

Couple of athletes' motion in contest is ruled by a Langevin type equation.
It is the Second Sacripanti's model (1989-90). It is possible to write, for the global motion

$$F = ma = -\mu v + u \sum_{j} (\pm 1)_j \delta(t - t_j) = F_a + F' \quad (11)$$

If the Couple of Athletes system moves by Brownian motion, it will be possible to find, the trajectory most probable and demonstrate that this probability satisfies the Fokker-Plank equation, which describes the variation of probability of Couple of Athletes presence on the Tatami, during time of competition

$$\frac{\partial f(q,t)}{\partial t} = -\frac{\partial}{\partial q}\left(Kf(q,t)\right) + \frac{1}{2}D\frac{\partial^2}{\partial q^2}f(q,t) \quad (12)$$

- K= - μq is the push –pull coefficient;
- D is the diffusion coefficient.
- Now remembering the Einstein relationship, the diffusion coefficient D can be correlated with the motion time evolution of the CM of Athletes couple.

In the limit of very short time interval or very long time interval, as regards to the square mean shift on the mat, which is connected to the energy; it is possible to write: for short times:

$$\langle x^2 \rangle \equiv \langle q^2 \rangle = 2Dt = 2\eta O_2 t^2 \quad (13)$$

and for very long times :

$$\langle x^2 \rangle \equiv \langle q^2 \rangle = 2Dt = 4\eta O_2 \frac{t}{\mu} \quad (14)$$

Then the diffusion coefficient D ( motion capability of couple) for very long times is proportional to the double of energy consumption and inversely proportional to friction coefficient, for very short times is proportional both to the energy consumption and to the time (this means that D in the last case it is not constant in time ).

If we consider not the motion of CM of couple but every single athlete it is a system of two masses connected by a spring (arms) the singular Langevin equations are:

$$\begin{aligned}
\cdot & \quad -\frac{kx_1}{m_1} - \frac{k}{2m_1}\left(\frac{\partial}{\partial x_1}(x_1 - x_2)^2\right) + \Gamma_1 \\
\cdot & \quad -\frac{kx_2}{m_2} - \frac{k}{2m_2}\left(\frac{\partial}{\partial x_2}(x_1 - x_2)^2\right) + \Gamma_2
\end{aligned} \quad (15)$$

If we assume that both random forces $\Gamma_2 \Gamma_1$ are not correlated ( each athlete fights against the other) the Fokker Plank equation, that describes the dynamic of couple, gives us the distribution function from which any average of macroscopic variable is obtained by integration:

$$\frac{\partial W}{\partial t} = \left\{\frac{\partial}{\partial x_1}v_1 + \frac{\partial}{\partial v_1}\left[\frac{kx_1}{m_1} - \frac{k}{2m_1}\frac{\partial(x_1 - x_2)^2}{\partial x_1} + \gamma_1 v_1\right] + \gamma_1 \frac{\kappa T}{m_1}\frac{\partial^2}{\partial v_1^2} - \frac{\partial}{\partial x_2}v_2 + \right\}$$

$$\left\{+\frac{\partial}{\partial v_2}\left[\frac{kx_2}{m_2} - \frac{k}{2m_2}\frac{\partial(x_1 - x_2)^2}{\partial x_2} + \gamma_2 v_2\right] + \gamma_2 \frac{\kappa T}{m_2}\frac{\partial^2}{\partial v_2^2}\right\}W \quad (16)$$

For example for the mean shifting velocity during the contest:

$$\langle v(t) \rangle = \int_{-\infty}^{+\infty} v(t) W(v,t) dv \quad (17)$$

Till now the only experimental solutions of this equation are evaluated by Japanese Researchers of the Kodokan Association in the far 1971:

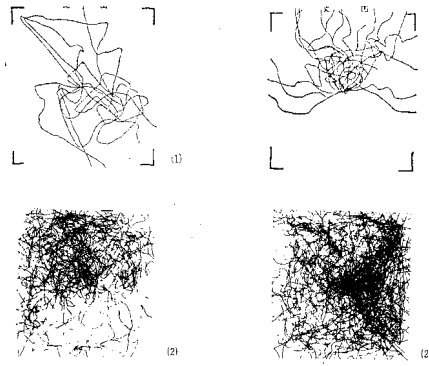

*Fig43. Judo contest shifting paths [9]*

1. The mean shifting velocity (**0.30 m/s**) during the contests [9].
2. The mean distance covered by judokas  (**121,1 m**) .

But the real motion is a most complicated 3D motion, how can we rebuild the most complex three-dimensional information, from the two dimensional path of Couple of Athletes motion?

*The projection theorem*
The problem previously proposed is that the motion path is a 2D projection of a 3D complex structured motion.
The assessment of volume properties from this 2D data is far from evidence, because few theoretical evaluation link 3D properties with 2D ones.
Only recently a very advanced theoretical study has shown that the H parameter of an n-dimensional isotropic fractal is linked to the (n-1)-dimensional fractals by the following self-similarity function:
$H_{(n-1)D} = H_{nD} + 0.5$  *that in our case is* $\rightarrow H_{2D} = H_{3D} + 0.5$  [28]
In the next figure is shown the meaning of the problem:

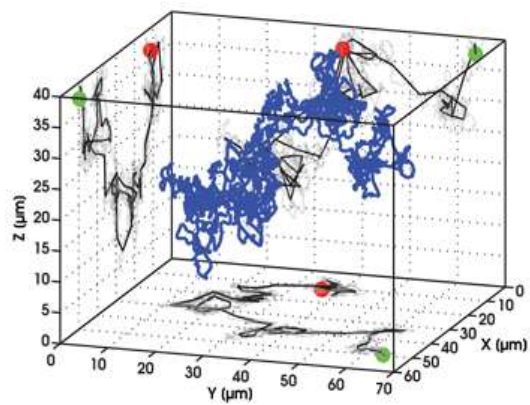

*Fig.44 Projection theorem for COM motion in Space*

*From the Brownian approach to the Newtonian ones*
Normally in real life not only the linear approximation but also the Newtonian Dynamics are in our common usage and knowledge.
How to relate Brownian dynamics and Newtonian mechanics?
The problem was yet solved by Einstein and other physics, in the study of diffusion.
We try to extent these results to shifing's paths in competition.
To describe the state of fight actually observed, it is convenient to introduce the 3D probability
f(**r**,**p**,t) d³**r** d³**p** to find Athlete's position **r** in the position-space volume d³**r** and momentum **p** in the momentum-space volume d³**p**, at time $t$.
Let us start from the sharp distribution

f(**r**,**p**,t= 0) = δ (**r** -**r**$_0$)δ(**p** -**p**)   (18)

Associated with the initial conditions of motion, where δ is the Dirac distribution.
We do not observe, at time $t > 0$, the sharp distribution associated with (18)

f(**r**,**p**,t) =  [**r** -**r**(t)]δ [**p** -**p**(t)],   (19)

Such as predicted by the Hamilton equations of the athlete's system

$$\frac{dr}{dt} = \frac{p}{m} \quad (20) \qquad \frac{dp}{dt} = F - \mu p \quad (21)$$

The solutions are easily evaluated:
$$p(t) = F\tau + (p_0 - F\tau)e^{-\mu t} \qquad (22)$$

$$r(t) = r_0 + \frac{F\tau}{m}t + \frac{\tau}{m}(p_0 - F\tau)\left[(1 - e^{-\mu t})\right] \qquad (23)$$

with $\tau = \frac{1}{\mu}$ the stationary solutions $t \gg \tau$ are

$$p(t) \approx \frac{F}{\mu} \qquad (24)$$

$$r(t) \approx r_0 + \frac{F}{\mu m}t \qquad (25)$$

To account for the sharp distribution in mechanical terms, Langevin added to the right-hand side of equation a stochastic force ξ $(t)$, also called *white noise.*
It is a random function of time with zero mean and a covariance proportional to the Kinetic energy produced, in our case." athletes fight", by the work of push-pull forces that are produced by third principle of dynamics between Athletes' bare feet and friction on the tatami.
For which: $\langle \xi(t) \rangle = 0$.
Working out Langevin's picture, it is found that *f* is the solution of the so-called Klein–Kramers equation,

$$\frac{\partial f}{\partial t} + \frac{p}{m}\frac{\partial f}{\partial r} + \frac{\partial}{\partial p}\left[\left(F - \mu p - \frac{\mu m v^2}{2}\frac{\partial}{\partial p}\right)f\right] = 0. \qquad (26)$$

Letting $E_k = \frac{\mu m v^2}{2} = 0$ in equation (26) gives a partial-differential equation on $f$ which, mathematically speaking, admits the Hamilton equations (20) and (21) as its characteristics.
Physically speaking, if $E_k = 0$, equation (26) admits solution (19) given the initial condition (18).
Now we have to do with two equivalent approaches spoken in two different languages.
Langevin's approach is cast in the language of dynamics with a novelty, namely the stochastic force $\xi(t)$.
When the latter is input into Newton's second law of motion, we obtain a stochastic differential equation.
In contrast, the approach of Klein and Kramers is expressed in statistical physics language.
The basic tool is a probability density in phase space, which is the solution of a linear partial-differential equation.
While Langevin's approach looks more intuitive at first glance and indeed was put forth before, its mathematics is subtle and open to criticism.
The approach of Klein and Kramers looks nebulous at first because of its use of phase space, but it is based upon standard mathematics.
We define the average of any observable $O$ from the probability density $f$ in the usual way of classical statistical mechanics:

$$\langle O \rangle \equiv \iint O f(r, p, t) d^3 r \, d^3 p \qquad (27)$$

It is possible to show by algebraic manipulations that equation (26) entails

$$\frac{d\langle r \rangle}{dt} = \frac{\langle p \rangle}{m} \qquad (28) \equiv (20)$$

$$\frac{d\langle p \rangle}{dt} = F - \mu \langle p \rangle \qquad (29) \equiv (21)$$

If we consider that F and μ are independent of position the Hamilton equations (20) and (21) are recovered on average.
Then the Newtonian approach is connected to the average on long time and space, while the fBm characterizes very short time and meso/micro space of observation.
Meso/microscopic and macroscopic trajectories are connected by the self affinity and theoretical self-similarity, also for anomalous diffusion H<0.5 and H>0.5 the scaling properties of these phenomena are connected sometime to the long term memory of the process.
Normally for persistent fight the correlation in time shows an inverse power law, normally 2 is greater of 2H, this means that correlation between two points in time decreases with increasing time separation, the spectrum (Fourier Transform of Correlation) is an inverse power law in frequency:

$$C(\tau) = \langle X(t) X(t + \tau) \rangle \propto t^{2H-2} \qquad (30)$$

$$S(\omega) = FT\{C(\tau), \omega\} \propto \frac{1}{\omega^{2H-1}} \qquad (31)$$

For the anti-persistent evolution of fight the spectrum, instead becomes a direct power law in frequency.
The evaluation, previously produced, is based on classical mechanics and classical statistical mechanics at meso /micro the fBm is the cornerstone of modeling.
Calculus of variation applied to the concept of ***Action*** is the most general and elegant form of classical mechanics.

The real world is a non conservative world, modeling the real world, by two different approaches, is similar in appearance but a deep analysis shows that the second one is very far from the previous one.
For conservative systems calculus of variation is equivalent to the method utilized by Newton
However while Newton allow non conservative forces, the techniques utilized by Lagrangian and Hamiltonian mechanics have no direct way to deal with them.
During the 1931 Bauer [29] demonstrated that it is impossible to use a variational principle in order to derive a non conservative linear equation of motion with constant coefficients.
Thirty years later was demonstrated by Riewe [30] that fractional derivative provide an elegant tool to solve this problem.
From that if Lagrangian is built by fractional derivative, the resulting equation of motion can be non conservative.
Then in real world friction is a ubiquitous presence and non conservative system is the normality, in such way to model the real situation clearly speaking is more complex than classical Newtonian, Hamiltonian approach and we need the help of fractional dynamics.

*Fractional approach: Real description of microscopic dimension*
The shifting patterns study, in the microscopic approach, is more complex but it could be source of very useful data, however the price to extract the hidden information is a non-trivial mathematical analysis of these special time series ( both length and time of stride).
The general principle of the fBm framework is that the aspect of a trajectory, expressed as a function of time, may be calculated by a fractional space dimension, hence providing a quantitative measurement of evenness of the trajectory.
It is possible to write in mathematical form:
$$D_t^a [X(t)] - \frac{X(0)}{\Gamma(1-\alpha)} t^{-\alpha} = \xi(t) \quad (32)$$
The first term is a fractional derivative, the second is connected to the initial condition of the process, and the third is always the random force acting on the COM.
The fractional Brownian motion has the following covariance :
$$\langle x(t_1) x(t_2) \rangle = D_H \left[ t_1^{2H} + t_2^{2H} - |t_1 - t_2|^{2H} \right] = \Gamma(1-2H) \frac{\cos \pi H}{2\pi H} \left[ t_1^{2H} + t_2^{2H} - |t_1 - t_2|^{2H} \right] \quad (33)$$

In this case is important to know the mean square displacement of the Athlete:
$$\langle [X(t) - X(0)]^2 \rangle = \frac{\langle \xi(t)^2 \rangle}{(2\alpha - 1) \Gamma(\alpha^2)} t^{2\alpha - 1} \propto t^{2H} \quad (34)$$

By this expression it is possible to understand that we are in presence of different spontaneous motions of Couple of Athletes System on the Tatami, which can be compared to different diffusion processes, identified by the Hurst parameter.
In particular the H parameter is time independent and it describes the fractional Brownian motion with anti-correlated samples for 0<H<1/2 and with correlated samples for ½<H<1.
If H is = to ½ we can speak of pure Brownian motion.
It is also very important that a fBm is connected to a Fractal based Poisson Point Process, this special feature will be very useful, in order to find a right theoretical basis to evaluate victory probability and short term forecasting in a Judo match.
Davidsen & Schuster [31] pay attention to a simple but plausible method for generating fractal-based point processes from ordinary Brownian motion.
Their construct resembles a conventional *integrate-and reset process*, but differs in that the threshold, rather than the integration rate, is taken to be a stochastic process.
This kind of behavior is really present in nature, it occurs in body's neurophysiology, for example, where ion-channel current fluctuations give rise to random threshold fluctuations.

In the model considered by Davidsen & Schuster the rate remains fixed and the threshold process is taken to be ordinary Brownian motion.
When the integrated state variable reaches the threshold, an output event is generated and the state variable is reset to a fixed value, as with a conventional integrated- reset process.
It is also important to see the autocorrelation coefficient of fBm that, as well known, depends only by the time increment.
The autocorrelation coefficient for all sorts of fBm depends only from the ratio τ/t where τ=t'-t

$$\rho(\tau,t) = \frac{1}{2}\left(\left|\frac{t}{\tau}\right|^{H} + \left|\frac{\tau}{t}\right|^{H} - \left|\sqrt{\left|\frac{t}{\tau}\right|} - \text{sgn}\left(\frac{\tau}{t}\right)\sqrt{\frac{\tau}{t}}\right|^{2H}\right) \quad (35)$$

For the special case ( τ = -t ) we have

$$\rho(\tau,t) = \rho(-t,t) = 1 - 2^{2H-1} \quad (36)$$

We remember also that only for H=1/2 (Regular Brownian Motion) autocorrelation coefficient for t and -t is independent, whereas fBm (t) and fBm(-t) are connected depending from the previous history. [31]

Athlete's Tracks (**Dromograms**) are the evolution in time of the couple of Athletes COM projection on the tratami area result of a spontaneous motion connected to the strategic thinking of Athletes .
Normally in the Match Analysis study, each technical action and throw is considered belonging to a class of Markovian System, this means that it depends by the previous instant only, without correlation with the past movements.
The more advanced mathematical approach based on Fractional Dynamics let able to overcome this conceptual limitation and mathematical simplification.
As we have seen before, an important feature of fBm modeling, for each fighter, is the presence of long-term correlations between past and future increments.
This means that the system is not Markovian and then more similar to real situation.
It is interesting to note that the human paths produced by strategic thinking are very similar to track produced by inanimate elements. This can be assessed by the scaling regimes.
In this way a fighting path can show, if correctly analyzed, when the fighter have a specific fighting strategy or not (kind of random motion) during competition.
For example a median value of 0.5 for *H* indicates that there is no correlation between actions, suggesting that the trajectory displayed a random distribution (Brownian motion) that is Markovian.
On the other hand, if *H* differs from 0.5, a positive (0.5 > *H)* or negative (H < 0.5) correlation with his fighting way can be inferred, indicating that a given part of initiative is under control.
In fact in the first case *H>0.5* the Fractional Brownian motion exhibits a long-range dependence; in contrast if *H<0.5* fractional Brownian motion is a short memory process.
Then depending on how *H* is positioned, with respect to the median value 0.5, it can be inferred that the subject more or less controls the trajectory (and the fight evolution in time).
The closer the regimes are to 0.5, the larger the contribution of stochastic processes (random actions without strategy). In addition, depending on whether *H* is greater or less than the 0.5 thresholds, persistent (attacking), anti-persistent (defending / counter striker) behavior can be revealed, respectively.
In other words, if the COM projection, at a certain time, is displaced towards a given direction, the larger probability is that it drifts away in this direction (persistent attacking behavior) or in the contrary it retraces its steps in the opposite direction (anti-persistent defensive / counter striker behavior).

The trajectory, obviously, contains more information than the mean squared displacement. In particular one can measure the waiting time distribution from stalling events in the trajectories.

For pronounced anti-persistent processes immobilization events should be observed, i.e., for certain time spans neither coordinate should show significant variation (athlete stops the shifting action). [32] Due to the scale-free nature of fBm anti-persistent these stops should span multiple time scales [33]. If such events occur they are indicative of the nature of the process very slow fighting with high deep study of the adversary.

Absence of such features in shorter time series cannot necessarily rule out the fBm dynamics, in particular for H closer to one (ballistic motion) distinct stops are relatively rare events and possibly require very long time and high intensity of fight.

Equality between these two probabilities ( $H= ½$) indicates that there is not presence of a defined strategy in fighting, like simple random motion or stochastic process.

This information obtained by a pure "mathematical lecture" of trajectories; can be enhanced adding to the previous mathematical lecture other complementary Biomechanical fighting information like: Competition Invariants, Action Invariants, Attack useful polygonal surface, Direction of displacement, Time and position of gripping action, Sen No Sen on grips, Throws "loci", Length or Amount of displacement, Mean Speed, Surface Area Utilization and so on.

It is possible, with this added complementary information, to obtain a lot of useful strategic information.

Information, ordered by importance or effectiveness, is useful for coaching and athletes as well.

This is one example of the advanced information obtainable by this underestimated practical tool: Athletes' shifting patterns.

Our analysis started from macroscopic Locomotion till to the microscopic approach to COM motion in 3D space [34] that is fBm, from that derives the connection to the fBm described by the perpendicular of Athlete COM on the mat, also without considering the standard error connected.

After that it is easy to induce that the motion of the Couple of Athletes COM perpendicular is again a fBm at microscopic level, but was also proved that, thanks to the self affinity, scaling at macroscopic level on long time interval and long space track, the motion is always Brownian.

In support of this connected reasoning a theoretical demonstration was presented into another paper [35].

In it is presented a physical Langevin-based theory (Newton connected, remembering that it is a formal derivation of a Langevin equation from classical mechanics), explaining the emergence and the pervasiveness of the 'fractional motions' like : Brownian motion, Levy motion, fractional Brownian motion and fractional Levy motion.

In the article a general form of "micro-level" Langevin dynamics, with infinite degrees of freedom, is presented. Scaling from the micro-level to the macro-level the many degrees of freedom are summarized in only two characteristic exponents: the Noah and the Joseph exponent and the aforementioned fractional motions emerge universally.

The previous two exponents categorize the fractional motions and determine their statistical and topological properties. This useful theory establishes unified 'Langevin bedrock' to fractional motions that as we know is the basic description of Judo shifting paths.

As shown Shifting's Trajectories study is a very hard and complex work, because it falls among fractal, self-similar trajectories produced by Chaotic (irregular) Dynamics.

Into such complex area is it possible to Optimize Athletes' motion planning in judo fight?

Optimization of stochastic systems is very complex and probably not solvable, but it is possible to overcome this difficulty changing the global vision of the system considering it among traffic problems.

Traditionally in traffic theory, motion planning is defined as the problem of finding a collision-free trajectory from the start configuration to the goal configuration.
It is possible to treat athletes'motion as an optimization problem, considering interaction between athletes and searching the trajectory that drive adversary into the final *broken symmetry position*, in which it is possible *to collide* and apply one of the two throwing tools (*Lever or Couple*).
In this totally new global vision it is easy to find for Coaches a Optimization solution.
In fact into this new field of study for Coaches: no matter how long is the path in the contest.
The important is to connect and study carefully *two or three steps that can bring to the goal configuration*, in which athlete after *symmetry break and collision* applies throwing tools *Lever or Couple* to throw the adversary.

## 9. Conclusions

This paper faces the problem of the Optimization of a Dual Situational Sport in is complex behavior , the solution found very useful for teachers and coaches, starts from the Kano solution Minimum of Energy consumption that is driven to a variational principle.
However this principle is strictly applicable only in still and simplified situations and then good for teaching purposes. For competition a new, more wide, variational principle must be used the "Principle of maximum effectiveness" .
This principle that includes the principle of minimum energy consumption, takes into consideration also technical actions that are energetically disadvantageous but very effective for the victory.
It is always in the area of variational calculus focalized on one extreme of the mechanical energy function.
However not all situations can be optimized in competition, during Interaction open skill evolves continuously in infinitive specific technical tactical solution.
More often, in such fast changing situations, for Coaches it is not possible plan before, optimization.
It is also introduced the notion of "Strategic Optimization" considering the Optimization of a whole judo contest or sequence that is called: *dynamic programming problem.*
*A dynamic programming problem* is an optimization problem in which decisions have to be taken sequentially over several time periods "linked" in some fashion.
 A *strategy* for a dynamic programming problem is just a *contingency plan.* i.e., a plan that specifies what is to be done at each stage as a function of all that has transpired up to that point.
 It is possible to demonstrate, under some conditions, that a Markovian optimal strategy is an optimal strategy for the *dynamic programming problem* under examination.
From that we build a Heuristic Energy Equation of the contest by a Constructive Algorithm that is an algorithm that constructs iteratively a final solution by building upon a partial (incomplete) solution.
This equation is again a variational problem considering the minimum of each contribution during the sequence of contests.
Finally a complete study of the hard problem of Couple of Athletes shifting paths is produced with many interesting points. However in term of Optimization only a short section of the complete shifting paths can be utilized to build a goal situation in which will be possible to collide and apply the throwing tools to produce a winning interaction

# 10. Appendix 1 Sphere and Cylinder Geodesics a "simple" calculus of variation.

If we like to find a geodesic for a sphere and for a right cylinder we must start from the general equation of the differential form of an arc on a figure:

$$(ds)^2 = (dr)^2 + (rd\theta)^2 + (dz)^2$$
$$\text{for a sphere} \quad r = \cos t \quad dr = 0 \quad dz = r\sin\theta\phi$$
$$\text{for a cylinder} \quad r = \cos t \quad dr = 0$$

$$\frac{ds}{d\theta} = c\sqrt{1 + \sin^2\theta\left(\frac{d\phi}{d\theta}\right)^2}$$

$$\frac{ds}{d\theta} = \sqrt{c^2 + \left(\frac{dz}{d\theta}\right)^2}$$

For both we must determine the minimum of s then the Euler equations are:

$$\frac{d}{d\theta}\left(\frac{\partial f}{\partial \phi}\right) = 0 \quad \frac{\partial f}{\partial \phi} = \cos t = k$$

$$\frac{d}{d\theta}\left(\frac{\partial f}{\partial z}\right) = 0 \quad \frac{\partial f}{\partial z} = \cos t = k$$

The solution of these two geodesics after some calculations problems are:

$$z = \left(\frac{\cos c_1\sqrt{1-k^2}}{k}\right)x + \left(\frac{\sin c_1\sqrt{1-k^2}}{k}\right)y$$

$$z = \left(\frac{k a\theta}{\sqrt{1-k^2}} + c_1\right)$$

The first is the plane that contain the maximum circle

The second is the circular Helix   [36]

# 11 Appendix 2 "Athletes shifting in computational Biomechanics"

# ( Prof. A. Pasculli  Chieti University) [37]

*Numerical Evaluation*
The aim of this evaluation is to formulate a numerical strategy to evaluate reasonable athlete's paths, linked, in a simplified way as a first step, to the previous discussed theory.
The fundamental assumptions are introduced and described through the following points:

a. the two athletes are supposed to be located at the two ends of a bar;
b. the bar can rotate around its middle point, considered as the barycentre of the *athletes couple*;
c. the length of the bar oscillates in a *harmonic-random* way;
d. The centre of the bar (barycentre of the couple) moves by a *random motion* like.

Thus, the motion of the two athletes is simulated by a *random motion* of a "*pulsating*", "*spinning*" *circle*, with a mass equal to the sum of the athletes' masses.
The variable diameter depends on both their gripping arms length and their shoulder "thickness".
In all the following numerical simulations, a total 140 kg weight and a 0.2-0.8 m diameter range are assumed.
From the previous discussion, it follows that the global motion of a single athlete is assumed to be the vectorial composition of three different elementary movements:
- A *random motion* of the centre of the "*athlete's couple*",
- A *random rotation* around the centre;
- A *random oscillation* towards the centre.
It is worth to observe that all the three elementary motions are characterized by a *random behaviour*.
The motion of the "*couple centre*" is characterized, at each step, by a direction *uniformly randomly chosen* (on 360 grades). Along the chosen direction a *displacement* is calculated supposing a *rectilinear uniformly accelerated* motion. The *time length* of the motion is evaluated, again, by a *random number generation approach*. In a more detailed way: the displacement of the *couple centre*, along the already chosen direction, is evaluated through the following equation ( Sacripanti):

$$m\frac{d\mathbf{v}}{dt} = -\beta\mathbf{v} - m\frac{(\mathbf{v}-\mathbf{v}_a)}{\tau} + \mathbf{P} - \mathbf{A}e^{-|x|/b} + \mathbf{L} \qquad (1)$$

Where m is the total mass of the athletes couple, $\mathbf{v}$ is the vector velocity, $\tau$ is the relaxation time necessary to reach the "target velocity" $\mathbf{v}_a$, $\mathbf{P}$ a *push/pull force*, $\mathbf{A}\exp(-|x|/b)$ is a global term related to the border line distance strategy, while $\mathbf{L}$ is the *random Langevin force*. This kind of force is introduced through the *random chosen of the direction* and the *random chosen* of how long is the "*couple centre*" displacement before the *next change of direction*. Thus the deterministic behaviour of this kind of motion is simulated by the solution of the following scalar equation:

$$m\frac{dv}{dt} = -\left(\beta+\frac{m}{\tau}\right)v + \left(\frac{mv_a}{\tau} + P - Ae^{-|x|/b}\right) \qquad (2)$$

Whose solution is easily?

$$\Delta s = \frac{c}{a}\Delta t + \left(\frac{v_0}{a} - \frac{c}{a^2}\right)\left(1 - e^{-a\Delta t}\right) \qquad (3)$$

Where $\Delta s$ is the total displacement of the "*couple centre*" displacement during the time step $\Delta t$,

$$a = \frac{\beta}{m} + \frac{1}{\tau}, \tag{4}$$

$$c = \frac{v_a}{\tau} + \frac{P}{m} - \frac{A}{m} e^{-|x|/b} \tag{5}$$

We assume the *time step is a random variable* $\Delta t_{rand}$ belonging to a Gaussian statistics with a mean time step $\Delta t_m$ and a variance $\sigma_{\Delta t}$.

Thus, we have considered the following expression for the selected time *random variable*:

$$\Delta t_{rand} = \Delta t_m + n_{\Delta t} \cdot \sigma_{\Delta t} \cdot G\_norm \tag{6}$$

Where $n_{\Delta t}$ indicates the total range of variability ($n_{\Delta t} = 2$ for all the following numerical simulation). Then a *normal* "G_norm" Gaussian distributed stochastic variable ($\mu = 0$ and $\sigma = 1$) can be provided by the *Box and Muller* (1958) algorithm:

$$G\_norm = \sqrt{[-2 \cdot \ln(Y1_{rand})]} \cdot \cos(2\pi \cdot Y2_{rand}) \tag{7}$$

Where $Y1_{rand}$ and $Y2_{rand}$ are two *independent* uniformly distributed random variables. That means that the intrinsic routine, related to the selected Compilator (RAND in FORTRAN 97), to generate random variables should be called. This subroutine *must be called twice* in order to obtain the two independent (pseudo-random) variables.
A simple uniform statistics is assumed as a first step, for a *random rotation* around the centre.
Thus, it was not implemented a more realistic dynamics introducing an *angular moment equation*.
The *random oscillation* towards the centre is simulated as a harmonic motion.
Future improvements will regard some reasonable assumptions in order to build an "objective function". The main criteria will be to select a function correlated to the *strategy* of the player around which, in a necessarily *randomly* way, a *tactic* function should be added. The *strategy* depends on the player characteristics.
Some numerical realizations of possible athletes paths, belonging to a statistical ensemble built up by the previous described approach have been carried up.
Fig. 1 shows how the athletes' movements are simulated by a *moving*, *spinning* and *pulsating* "*athletes couple circles*", shown in the red colour. The athletes are located at the ends of the reported red diameters, while the tracks of one of the two athletes are reported in black lines.

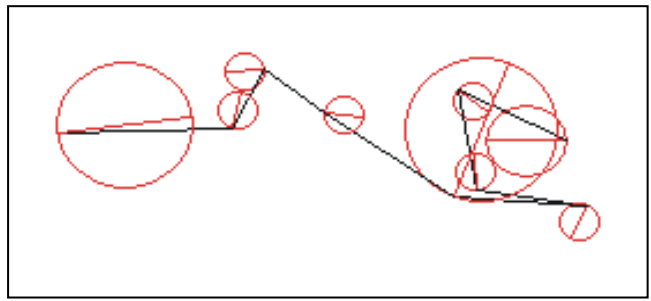

Fig. 1 Movements of the "*athlete's couple*"

In Fig.2 a single game, related to a possible realization belonging to a supposed statistical ensemble, is reported.

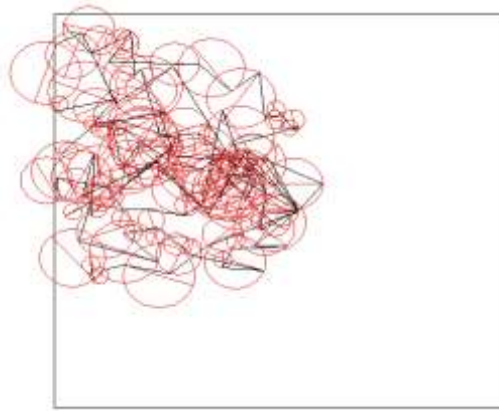

*Fig. 2 Tracks and "athletes couple circles"*

*A single game "realization"*

Fig.3 shows a superposition *realization* of the tracks of a single athlete, belonging to the *same statistical ensemble*, regarding *16 different virtual games*. The ensemble are built up supposing a completely symmetry.

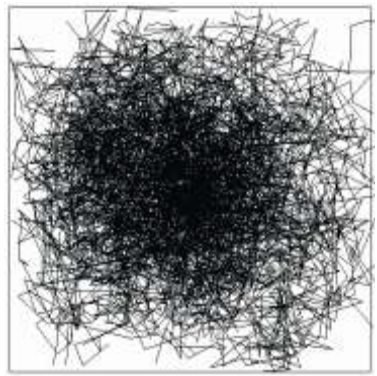

*Fig.3 Realizations of 16 virtual games*

In Fig.4a and 4b two different realizations of the tracks of a single athlete, belonging to the same statistical ensemble, regarding a single game, are reported.
The ensemble is built up supposing a smooth asymmetry along N-W direction.
Fig. 4c shows recorded tracks of an actual single game.
Also in Fig.5a and 5b two others different realizations of the tracks of a single athlete, belonging to the same statistical ensemble, regarding a single game, are reported.
This new ensemble is built up supposing a smooth asymmetry along S-N direction.
Fig. 5c shows recorded tracks of an actual single game.
In Figs. 6a, 6b, 6c; and Figs. 7a, 7b, 7c a superposition of both 7 and 12 single tracks, compared with a superposition of actual games are reported.
It is important to note that the changes of some parameters described before, allow describing different actual tracks. Another important point to note is the occurrence of *spatial permanence* of the tracks (Figs. 4a, 4b, 4c in particular). As the tracks depend on random movements, it seems that the *spatial permanence* could be related to the *Fractional Brownian* like motion (fBm) of the tracks (actual tracks as well).

| Fig. 4a N-W asymmetry | 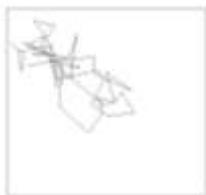 | 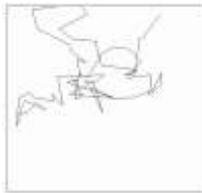 | Fig. 5a S-N asymmetry |
| --- | --- | --- | --- |
| Fig. 4b N-W asymmetry | 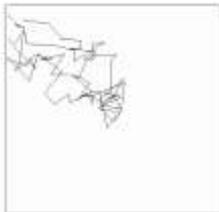 | 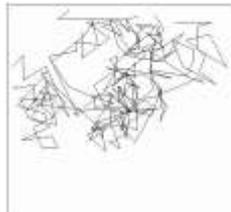 | Fig. 5b S-N asymmetry |
| Fig. 4c One game | 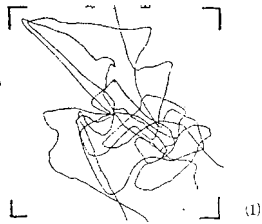 | 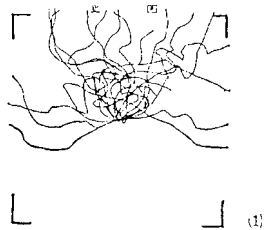 | Fig. 5c One game |
| Fig. 6a 7 games | 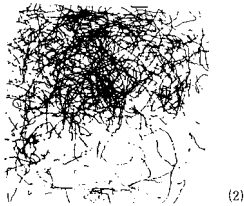 | 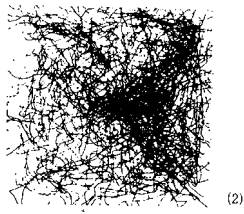 | Fig. 7a 12 games |
| Fig. 6b N-W asymmetry – | 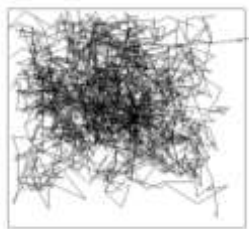 | 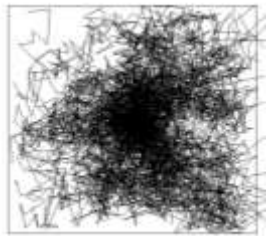 | Fig. 7b N-E asymmetry |
| Fig. 6c N-W asymmetry – | 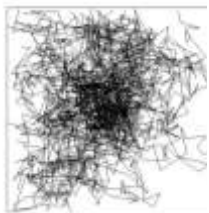 | 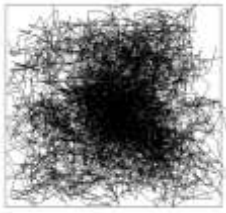 | Fig. 7c N-E asymmetry |